\documentclass[12pt,prd,superscriptaddress,nofootinbib]{revtex4}
\usepackage[english]{babel}
\usepackage{graphicx}
\usepackage{mathtext}
\usepackage{indentfirst}
\usepackage{epsfig,amsmath,amsfonts}
\newcommand{\beq}[1]{\begin{eqnarray}\label{#1}}
\newcommand\eeq {\end{eqnarray}}
\newcommand\bqa {\begin{eqnarray}}
\newcommand\eqa {\end{eqnarray}}
\newcommand\pr {\partial}

\newcommand{\bear}{\begin{array}}
\newcommand{\enar}{\end{array}}

\begin{document}

~~~~~~~~~~~~~~~~~~~~~~~~~~~~~~~~~~~~~~~~~~
\hfill ITEP--TH--2/12

\hfill AEI--2012--004

\vspace{5mm}

\title{\Large Comparative study of loop contributions\\
in AdS and dS}
\author{\bf E.\ T.\ Akhmedov}
%\email{akhmedov@itep.ru}
\affiliation{B.\ Cheremushkinskaya, 25, ITEP, 117218, Moscow, Russia}
\affiliation{Moscow Institute of Physics and Technology, Dolgoprudny, Russia}
\affiliation{Max-Planck-Institut f\"ur Gravitationsphysik (Albert-Einstein-Institut),
Am M\"uhlenberg 1, 14476 Golm, Germany}
\author{\bf A.\ V.\ Sadofyev}
%\email{sadofyev@itep.ru}
\affiliation{B.\ Cheremushkinskaya, 25, ITEP, 117218, Moscow, Russia}
\affiliation{Moscow Institute of Physics and Technology, Dolgoprudny, Russia}

%\centerline{\Large \bf Comparative study of loop contributions in AdS and dS}

%\vspace{10mm}

%\centerline{\bf E.T.Akhmedov and A.V. Sadofyev}

%\vspace{5mm}

%\centerline{Institute of Theoretical and Experimental Physics, Moscow}

%\centerline{Moscow Institute of Physics and Technology, Dolgoprudny}

%\date{}

\maketitle

\begin{center}{\bf Abstract}\end{center}
The generic feature of non--conformal fields in Poincare patch of de Sitter space is the presence of large IR loop corrections even for massive fields. Moreover, in global de Sitter there are loop IR divergences for the massive fields. Naive analytic continuation from de Sitter to Anti--de--Sitter might lead one to conclude that something similar should happen in the latter space as well. However, we show that there are no large IR effects in the one--loop two--point functions in the Poincare patch of Anti--de--Sitter space even for the zero mass minimally coupled scalar fields. As well there are neither large IR effects nor IR divergences in global Anti--de--Sitter space even for the zero mass.

\vspace{5mm}

\section{Introduction}

There are large IR loop effects in de Sitter (dS) space. E.g. consider two--point Keldysh function in say massive, minimally coupled, scalar field theory, $G^K(\eta_1, \vec{x}_1; \eta_2, \vec{x}_2) \equiv \frac12\, \left\langle \left\{\phi\left(\eta_1,\vec{x}_1\right),\, \phi\left(\eta_2,\vec{x}_2\right)\right\}\right\rangle$, where $\{\cdot, \cdot\}$ means anti--commutator. Here the coordinates on the $d$--dimensional dS space of unit curvature are defined as $ds^2 = (d\eta^2 - d\vec{x}^2)/\eta^2$, the limit $\eta \to + \infty$ of the conformal time $\eta$ corresponds to the past, while $\eta \to 0$ corresponds to the future in the expanding Poincare patch of dS. Poincare patch covers half of dS space.

Due to spatial homogeneity of dS space it is convenient to consider the spacial Fourier transform of this function: $G^K(\eta_1, \eta_2;\vec{p}) = \int d^{d-1}x \, e^{i\, \vec{p}\, \vec{x}}\, G^K(\eta_1, \vec{x}; \eta_2, 0)$. The IR contributions to the exact Keldysh propagator, in the limit $p\sqrt{\eta_1\eta_2}\to 0$ and $\eta_1/\eta_2=const$, can be represented as follows \cite{Krotov:2010ma}\cite{Akhmedov:2011pj}:

\bqa\label{eq1}
G^K \left(\eta_1, \eta_2; \vec{p}\right) = \left(\eta_1\eta_2\right)^{\frac{d-3}{2}} \,\Biggl\{\frac12\, \biggl[h(p\eta_1)\, h^*(p\eta_2) + h^*(p\eta_1) \, h(p\eta_2)\biggr]\, \biggl[1 + 2\, n_p\left(\eta\right)\biggr] \Biggr. + \nonumber \\ + \Biggl. h(p\eta_1)\, h(p\eta_2)\,\kappa_p\left(\eta\right) + h^*(p\eta_1)\, h^*(p\eta_2)\,\kappa^*_p\left(\eta\right)\Biggr\}
\eqa
for the scalars with $m>(d-1)/2$. The calculation should be done in the non--stationary Keldish--Schwinger diagrammatic technic. Here $\eta = \sqrt{\eta_1\, \eta_2}$ is the average conformal time, $h(p\eta)$ is a solution of the Bessel equation with the index $\mu_{dS}^2 = m^2 - \left(\frac{d-1}{2}\right)^2$ and $m$ is the mass of the particle. The choice of $h(p\eta)$ defines the background state. E.g. the Bunch--Davies \cite{Bunch:1978yq} state corresponds to the Hankel function in place of $h(p\eta)$.
If one starts from the vacuum state,
then at tree level $n_p = \langle a^+_p \, a_p \rangle = 0$ and $\kappa_p = \langle a_p \, a_{-p}\rangle = 0$.

Due to the particle creation $n_p$ and $\kappa_p$ are not zero at the loop level.
E.g. for Bunch--Davies initial state in the scalar field theory with the $\lambda \, \phi^3$ self interaction one obtains that at one--loop level $n_p(\eta) \propto \lambda^2 \, \log(p\eta/\mu_{dS})$ and $\kappa_p(\eta) \propto \lambda^2 \, \log(p\eta/\mu_{dS})$ \cite{Krotov:2010ma},\cite{Akhmedov:2011pj},\cite{Leblond}. For the $\alpha$--vacua the situation is similar with the only difference that for the vacuum defined by the standard Bessel functions (out--Jost harmonics)
$n_p(\eta) \propto \lambda^2 \, \log(p\eta/\mu_{dS})$, while $\kappa_p(\eta)$ is suppressed \cite{Akhmedov:2011pj}.

Similar IR contributions (with different powers of the coupling constants and of the $\log(p\eta)$) do appear in different types of correlation functions and for the scalar fields with the other types of self interactions and masses. The same is true for the other non--conformal theories in dS space. This is the generic behavior in the Poincare patch of dS \cite{Woodard}, \cite{Dolgov:1994cq}, \cite{Antoniadis:2006wq}, \cite{Xue:2012wi}, \cite{Giddings:2010ui}.

The situation becomes even more interesting in the global dS space. There the loop contributions are explicitly IR divergent \cite{Akhmedov:2008pu}, \cite{Krotov:2010ma} even for the massive fields. This fact makes dS space similar to the QED in strong background electric fields \cite{Akhmedov:2009vh}. The presence of the IR cut--off in the correlation functions puts an abstraction for the dS isometry invariance of the correlation functions of QFT in global dS. That could lead to the dynamical quantum secular screening of the cosmological constant.

The immediate question is whether such an IR behavior is the specific feature of the quantum fields in dS space or there is a similar behavior of the fields on the other symmetric curved space --- on Anti--de--Sitter (AdS) space? Indeed the naive analytic continuation from dS to AdS might lead one to conclude that similar IR effects should appear in AdS as well. In this note we show that such an expectation is wrong.

Covering space of AdS has preferred reference frame which corresponds to inertial observers, covers entire space--time and is static. Hence, QFT in such a background has unique AdS invariant vacuum state. In this respect it is similar to the Minkowski space and is different from the dS space. As the result one can expect that IR effects in EAdS should not be too strong. Furthermore, one can study QFT dynamics on the covering of AdS space via straightforward analytic continuation of the Feyman diagrammatic expressions from Euclidian AdS (EAdS) space.

We show that there are no large IR effects in the one--loop two--point functions in the Poincare patch of (E)AdS space even for zero mass scalar fields.
As well there are neither large IR effects nor IR divergencies in global (E)AdS space. The methods we use in this paper are similar to those, which were used in \cite{Marolf:2010zp} for the sphere.

\section{General discussion of scalar fields in EAdS}

The $d$--dimensional EAdS space is the hyperboloid $X^2 \equiv X_0^2 - X_i^2 = 1$ in the flat $(d+1)$--dimensional Minkowski space $ds^2_{d+1} = dX_0^2 - dX_i^2$. We set its curvature to 1 throughout this note.
Poincare patch covers only half of EAdS. In fact, the induced coordinates on this patch are defined as

\bqa
X_0 = \frac12\, \left(z + \frac{1}{z}\right) + \frac{x_a^2}{2\, z}, \quad X_d = \frac12\, \left(z - \frac{1}{z}\right) + \frac{x_a^2}{2\, z}, \nonumber \\
X_a = \frac{x_a}{z}, \quad a=1,\dots, d-1,
\eqa
and for this choice of coordinates $X_0 - X_d \ge 0$, because $z>0$.
Then the induced metric and Laplacian on such a $d$--dimensional space is as follows

\begin{eqnarray}
ds^2=\frac{dz^2+dx_a^2}{z^2}, \quad {\rm and} \quad
\triangle = z^2\,\partial_z^2+(2-d)\,z\,\partial_z+ z^2\,\triangle_{d-1},
\end{eqnarray}
where $\triangle_{d-1}$ is the flat $(d-1)$--dimensional Laplacian.

Global coordinates on the $d$--dimensional EAdS are induced via $X_0 = \cosh\rho$ and $X_i = \omega_i \, \sinh\rho$, $i=1,\dots,d$,
where $ \omega_i^2 = 1$ defines $(d-1)$--dimensional sphere. Then the induced metric and the Laplacian on the space in question is:

\bqa
ds^2 = d\rho^2 + \sinh^2 \rho \, d\Omega^2,\quad {\rm and} \quad
\Delta = \pr^2_\rho + (d-1)\, \coth\rho \, \pr_\rho + \frac{\Delta_{\Omega}}{\sinh^2\rho},
\eqa
where $d\Omega^2$ and $\Delta_\Omega$ are the metric and Laplacian on the $(d-1)$--dimensional sphere.

The Feynman propagator $G(X_1,X_2)$ of the minimally coupled massive, $m$, scalar QFT in EAdS should be a solution of the Klein--Gordon equation. This equation is invariant under the EAdS isometry. Hence, its solution should be a function of the EAdS invariant combination of the two points $X_1$ and $X_2$ \cite{Burgess:1984ti}. Such a combination is the hyperbolic distance $Z_{12} = X_1\cdot X_2$, where $X_1$ and $X_2$ are the coordinates in the ambient space of the two points on the hyperboloid, i.e. $X^2_1 = X^2_2 = 1$. In the Poincare coordinates the hyperbolic distance is $Z_{12} = 1+\frac{(z_1-z_2)^2+\left|\vec{x}_{1} - \vec{x}_{2}\right|^2}{2z_1z_2}$, while in global EAdS $Z_{12} = \cosh(\rho_1) \, \cosh(\rho_2) - \sinh(\rho_1)\, \sinh(\rho_2) \, \cos\Delta\varphi$, where $\Delta\varphi$ is the angle between the two vectors on the spherical $\rho$--sections.
In EAdS the hyperbolic distance ranges as $1\leq Z_{12}<\infty$.

The Klein--Gordon operator when acting on the function of the hyperbolic distance $Z$, rather than the function of the two points $X_1$ and $X_2$ separately, is equivalent to (see e.g. \cite{Allen:1985ux},\cite{Mottola:1984ar} for a similar discussion in dS space)

\bqa
-\Delta + m^2 = \left(1 - Z^2\right)\, \pr_Z^2 - d\, Z\, \pr_Z + m^2,
\eqa
which converts the Klein--Gordon equation into a form of the hypergeometric equation.
The solution of the latter equation, which is finite as $Z\rightarrow\infty$ is as follows

\begin{eqnarray}
\label{prop}
G(Z)\propto Z^{-\mu-\frac{d-1}{2}}F\left(\frac{d-1}{4} + \frac{\mu}{2},\frac{d+1}{4}+\frac{\mu}{2};\mu+1;\frac{1}{Z^2} + i\epsilon\right),
\end{eqnarray}
where $\mu^2=\left(\frac{d-1}{2}\right)^2+m^2$ and $F$ is the${\phantom{1}}_2F_1$ hypergeometric function. This propagator corresponds to the unique EAdS invariant vacuum state \cite{Burgess:1984ti} and is related via analytical continuation to the in--out (Q--)propagator in dS space \cite{Polyakov:2007mm},\cite{Akhmedov:2009ta}. The presence of $i\epsilon$ in the argument of the hypergeometric function is the usual shift, $Z\to Z-i\epsilon$, of the pole at $Z=1+i\epsilon$, corresponding to the coincident points. Note that the propagator in question has as well the pole at $Z=-1+i\epsilon$ on the complex $Z$--plane. It is an analytical function on this plane with the cut going from $Z=-1+i\epsilon$ to infinity just above the real axis.

Throughout this paper we are going to study one--loop contribution in the minimally coupled massive real scalar field theory:

\bqa
L = \sqrt{|g|}\,\left[\frac{g^{\mu\nu}}{2}\, \pr_\mu \phi \, \pr_\nu\phi + \frac{m^2}{2}\, \phi^2 + \frac{\lambda}{3}\, \phi^3 + \dots\right].
\eqa
Dots here stand for the higher self--interaction terms, which make the theory stable.
The reason why we are going to consider below formulas only due to the unstable cubic part of the potential is just to simplify them. This instability does not affect our conclusions.

The main subject of this note is the IR behavior of the one--loop contribution. But let us say a few words about the UV divergencies. Obviously they should be the same as in flat or dS space, because short wavelength fluctuations of the fields are not sensitive to the large curvature of the space--times.

In the UV limit $\left|X_1 - X_2\right|\to 0$ and, hence, $\frac{(z_1-z_2)^2+|\vec{x}_{1} - \vec{x}_{2}|^2}{2z_1z_2}\ll 1$. In this limit 
the propagator acquires the form: $G(Z)\propto 1/(Z-1)^{\frac{d-1}{2}}$.
The one--loop contribution to the propagator is

\begin{eqnarray}
\label{loop}
G_{1loop}(1,4)=\lambda^2 \, \int G\bigl(Z_{12}\bigr)\,G^2\bigl(Z_{23}\bigr)\,G\bigl(Z_{34}\bigr) \frac{dz_2dV_2}{z_2^d} \frac{dz_3dV_3}{z_3^d},
\end{eqnarray}
where $V$ is $(d-1)$--dimensional flat spatial volume. The leading UV divergence comes from:

\begin{eqnarray}
G^{UV}_{1loop}(1,4)\propto\lambda^2 \, \int G(Z_{12})\frac{1}{(Z_{23}-1)^{d-1}}G(Z_{34}) \frac{dz_2dV_2}{z_2^d} \frac{dz_3dV_3}{z_3^d}.
\end{eqnarray}
The integration variables $X_2,X_3$ can be changed to $\rho=\frac{X_2+X_3}{2},\epsilon=\frac{X_2-X_3}{2}$. In the UV limit $|\epsilon|$ is negligible with respect to $|\rho|$. Then one can approximate $Z_{23}-1\simeq \epsilon^2/2\rho_z^2~,~z_2z_3\simeq\rho_z^2~,~Z_{12}\simeq Z_{1\rho}~,~Z_{34}\simeq Z_{\rho4}$, where $\epsilon^2=\epsilon_z^2+\epsilon_i^2$. As the result we obtain

\begin{eqnarray}
G^{UV}_{1loop}(1,4)\propto\int d^d\rho \frac{G(Z_{1\rho})G(Z_{\rho4})}{\rho_z^{4}}\int_0  \frac{d\epsilon}{\epsilon^{d-3}}.
\end{eqnarray}
Thus, as expected, the UV divergence in EAdS space is the same as in flat and dS spaces.

\section{Poincare patch of EAdS}

In this section we consider the IR behavior of the one--loop contribution to the scalar field propagator in the Poincare patch of EAdS space. To do the one loop calculation in a curved space one in general has to specify the coordinate system, because the result of the calculation can depend on the choice of the coordinates if there are large IR effects. In fact, in different coordinate systems one, in principle, can specify different types of boundary conditions, which is usually done implicitly (see e.g. \cite{Akhmedov:2011pj} for a similar discussion in dS). The boundary conditions are necessary in those situations when coordinates under consideration cover only some part of the whole space. Large IR effects, if any, are obviously sensitive to the boundary conditions.

Due to the spatial homogeneity of EAdS space we find it more convenient to perform the Fourier transformation of the propagator (\ref{prop}) over the $d-1$ spatial dimensions.
The eigenfunctions of the Klein--Gordon operator are $z^{\frac{d-1}{2}}K_\nu(kz)e^{i\,\vec{k}\,\vec{x}}$ and correspond to the eigenvalues $\nu^2 - \mu^2$, where $K$ is the MacDonald function. Then the basis of the so called Kontorovich--Lebedev integral transformation is the following resolution of the $\delta$--function:

\begin{eqnarray}
\delta(x-y) = \frac{1}{\sqrt{x\,y}}\,\int_{0}^\infty d\tau\,\frac{2\tau\sinh{\pi\tau}}{\pi^2} \, K_{i\tau}(x)\, K_{i\tau}(y).
\end{eqnarray}
Using it, one can represent the Fourier transform of (\ref{prop}) as

\begin{eqnarray}
G(z_1,z_2,k)=-\frac{1}{\pi^2}(z_1z_2)^{\frac{d-1}{2}}\int_{0}^{\infty}\frac{2\,\tau\,\sinh{\pi\tau}}{\tau^2+\mu^2}K_{i\tau}(kz_1)K_{i\tau}(kz_2)d\tau.
\end{eqnarray}
The $\tau$ integration can be done using 2.16.52.11, 2.16.52.12 of \cite{Integrals}:

\begin{eqnarray}
\label{Fourier}
G(z_1,z_2,k) = \frac{\pi^2}{2}(z_1z_2)^{\frac{d-1}{2}}\, \left[I_\mu(kz_1)\,K_\mu(kz_2)^{\phantom{\frac12}}\theta(z_2-z_1) + I_\mu(kz_2)\,K_\mu(kz_1)^{\phantom{\frac12}}\theta(z_1-z_2) \right].
\end{eqnarray}
Here $I$ is another MacDonald function.
It is not hard to see, using 2.16.31.2, 2.16.31.3 of \cite{Integrals}, that the inverse Fourier transform of the last expression is (\ref{prop}), if $d=4$. In the other dimensions the situation should be similar.

 The one--loop contribution (\ref{loop}) in the Fourier transformed form is as follows:

\begin{eqnarray}
G_{1loop}(z_1,z_4,p)=\lambda^2 \, \int G(z_1,z_2,p)G(z_2,z_3,q)G\left(z_2,z_3,|p-q|\right)G(z_3,z_4,p) \frac{d^{d-1}q}{\left(2\,\pi\right)^{d-1}}\frac{dz_2dz_3}{z_2^dz_3^d}.
\end{eqnarray}
Substituting (\ref{Fourier}) into this expression and
making the change of variables $x_i = qz_i, \,\, i=1,2,3,4$ in the last expression, we obtain that the one--loop contribution contains the following integrals:

\begin{eqnarray}
\label{nz1}
\int_{x_1}^\infty dx_2\int_0^{(x_2,x_4)} dx_3 (x_2x_3)^{\frac{d-3}{2}}I_\mu\left(\frac{p}{q}x_3\right)K_\mu\left(\frac{p}{q}x_2\right) I_\mu(x_3)K_\mu(x_2)I_\mu\left(\frac{|p-q|}{q}x_3\right)K_\mu\left(\frac{|p-q|}{q}x_2\right),
\end{eqnarray}

\begin{eqnarray}
\label{nz2}
\int_{x_1}^{x_4} dx_3\int_{x_1}^{x_3} dx_2 (x_2x_3)^{\frac{d-3}{2}}I_\mu\left(\frac{p}{q}x_3\right)K_\mu\left(\frac{p}{q}x_2\right) I_\mu(x_2)K_\mu(x_3)I_\mu\left(\frac{|p-q|}{q}x_2\right)K_\mu\left(\frac{|p-q|}{q}x_3\right),
\end{eqnarray}

\begin{eqnarray}
\label{nz3}
\int_{x_1}^\infty dx_2\int_{x_4}^{x_2} dx_3 (x_2x_3)^{\frac{d-3}{2}}K_\mu\left(\frac{p}{q}x_3\right)K_\mu\left(\frac{p}{q}x_2\right) I_\mu(x_3)K_\mu(x_2)I_\mu\left(\frac{|p-q|}{q}x_3\right)K_\mu\left(\frac{|p-q|}{q}x_2\right),
\end{eqnarray}

\begin{eqnarray}
\label{nz4}
\int_{x_4}^\infty dx_3\int_{x_1}^{x_3} dx_2 (x_2x_3)^{\frac{d-3}{2}}K_\mu\left(\frac{p}{q}x_3\right)K_\mu\left(\frac{p}{q}x_2\right) I_\mu(x_2)K_\mu(x_3)I_\mu\left(\frac{|p-q|}{q}x_2\right)K_\mu\left(\frac{|p-q|}{q}x_3\right),
\end{eqnarray}

\begin{eqnarray}
\label{nz5}
\int_{x_4}^{x_1} dx_2\int_{x_4}^{x_2} dx_3 (x_2x_3)^{\frac{d-3}{2}}I_\mu\left(\frac{p}{q}x_2\right)K_\mu\left(\frac{p}{q}x_3\right) I_\mu(x_3)K_\mu(x_2)I_\mu\left(\frac{|p-q|}{q}x_3\right)K_\mu\left(\frac{|p-q|}{q}x_2\right),
\end{eqnarray}

\begin{eqnarray}
\label{nz6}
\int_{x_4}^\infty dx_3\int_0^{(x_1,x_3)} dx_2 (x_2x_3)^{\frac{d-3}{2}}I_\mu\left(\frac{p}{q}x_2\right)K_\mu\left(\frac{p}{q}x_3\right) I_\mu(x_2)K_\mu(x_3)I_\mu\left(\frac{|p-q|}{q}x_2\right)K_\mu\left(\frac{|p-q|}{q}x_3\right),
\end{eqnarray}

\begin{eqnarray}
\label{nz7}
\int_0^{x_1} dx_2\int_0^{(x_2,x_4)} dx_3 (x_2x_3)^{\frac{d-3}{2}}I_\mu\left(\frac{p}{q}x_2\right)I_\mu\left(\frac{p}{q}x_3\right) I_\mu(x_3)K_\mu(x_2)I_\mu\left(\frac{|p-q|}{q}x_3\right)K_\mu\left(\frac{|p-q|}{q}x_2\right),
\end{eqnarray}

\begin{eqnarray}
\label{nz8}
\int_0^{x_4} dx_3\int_0^{(x_3,x_1)} dx_2 (x_2x_3)^{\frac{d-3}{2}}I_\mu\left(\frac{p}{q}x_2\right)I_\mu\left(\frac{p}{q}x_3\right) I_\mu(x_2)K_\mu(x_3)I_\mu\left(\frac{|p-q|}{q}x_2\right)K_\mu\left(\frac{|p-q|}{q}x_3\right),
\end{eqnarray}
and is defined as follows

\begin{eqnarray}\label{expr}
G_{1loop}(z_1,z_4,p) = \lambda^2 \, \left(z_1\,z_4\right)^{\frac{d-1}{2}} \times \nonumber \\ \int_0^{+\infty}\frac{dq}{q}\left\{I_\mu(pz_1)K_\mu(pz_4)^{\phantom{\frac12}}\biggl[(\ref{nz1}) + (\ref{nz2})\biggr] + I_\mu(pz_4)K_\mu(pz_1)\,\biggl[(\ref{nz5}) + (\ref{nz6})\biggr] + \right. \nonumber \\ +\left. I_\mu(pz_1)I_\mu(pz_4)\,\biggl[(\ref{nz3}) + (\ref{nz4})\biggr]+K_\mu(pz_1)K_\mu(pz_4)^{\phantom{\frac12}}\biggl[(\ref{nz7}) + (\ref{nz8})\biggr]\right\}.
\end{eqnarray}
The expressions $(x_i,x_j)$ as the limits of integration in these formulas denote that the integration is going to that value among $x_i$ and $x_j$ which is smaller.

As the side remark let us point out that one can show that the obtained one--loop result is the function of $Z$, i.e. is invariant under EAdS isometry. The question appears because Poincare patch covers only half of EAdS and, hence, brakes the isometry. In particular, in the loop integrals this fact reveals itself via the non--invariance of the measure of integration in the vertices.
The measure is defined in terms of the coordinates $X_0,\dots, X_d$ in the ambient space as $d^{d+1}X \, \delta\left(X^2 - 1\right)\, \theta\left(X_0 - X_d\right)$. (This measure is equivalent to $\frac{dz}{z^d}\, d^{d-1}x$.) The presence of the Heaviside $\theta$--function, restricting to the Poincare patch, brakes the EAdS isometry. The latter is just the rotation group of the ambient $(d+1)$--dimensional space. However, one can show that under the infinitesimal rotation around $X_0$ towards say $X_1$ ($X_d \to X_d - \varphi \, X_1$) the one loop contribution changes by the integral over the measure $d^{d+1}X \, \delta\left(X^2-1\right)\, \varphi \, X_1 \, \delta(X_0-X_d) = d(X_0+X_d)\, d^{d-1}X \delta\left(X^2 - 1\right)\, \varphi \, X_1$. This integral does vanish because its integrand is the product of the Feynman propagators in EAdS (\ref{prop}), which, due to the shift $Z\to Z-i\epsilon$, are the analytical functions of $X_0+X_d$ in the lower half complex plane\footnote{We would like to thank A.Polyakov for telling us the idea of this proof. The same observation was used to prove the dS isometry invariance of the one--loop contribution in Poincare patch over the Bunch--Davies vacuum in scalar field theory. (Some elements of the proof can be found in \cite{Polyakov:2007mm},\cite{Polyakov:2009nq}.) At the same time one can show that one--loop contributions for the $\alpha$--vacua are not dS invariant, because propagators in this case have different analytical properties on the complex $Z$ plane from those of Bunch--Davies propagator. The question which vacuum in dS space is stable under the small density perturbations was addressed in \cite{Akhmedov:2011pj}. It follows from the kinetic equation \cite{Akhmedov:2011pj} that Bunch--Davies vacuum is unstable under the small perturbation, while the stable one is defined with respct to the out--Jost harmonics.} and decay at the complex infinity as a power of $Z$.

Returning back to the IR behavior, note that in the Poincare patch of dS space the one--loop contribution to the two--point function has the divergence in the limit $p\, \sqrt{\eta_1\eta_2}\to 0$, $\eta_1/\eta_2=const$. Similar limit in EAdS is $p\sqrt{z_1 z_4}\to 0$, $z_1/z_4=const$ and via analytical continuation one may expect that in EAdS the one--loop contribution behaves as $\log(p\sqrt{\eta_1\eta_2}/\mu_{dS}) \to \log(p\sqrt{z_1\,z_4}/\mu)$ in the limit in question. However this expectation is wrong.

In fact, using the behavior of the MacDonald functions $K_\mu(x)\simeq\frac{\Gamma(\mu)}{2}\left(\frac{2}{x}\right)^\mu$, $I_\mu(x)\simeq\frac{1}{\Gamma(\mu+1)}\left(\frac{x}{2}\right)^\mu$, as $x\to 0$, it is straightforward to show that (\ref{expr}) behaves as $p^{d-1}$ in the described limit, i.e. is not divergent. That is true even for the massless fields. So in Poincare patch of EAdS there are no large IR effects even for the massless scalars.

\section{Global EAdS}

We go on with the global EAdS calculation. The eigen--functions of the Klein--Gordon operator in the global EAdS are $H_{p,l,m}\left(\rho, \Omega\right) = Z_{p,l}(\rho) \, Y_{l,m}\left(\Omega\right)$ \cite{Grosche:1987de}, where $Y_{l,m}\left(\Omega\right)$ are spherical harmonics on the $(d-1)$--dimensional sphere: $\Delta_{\Omega} Y_{l,m}\left(\Omega\right) = - l\,(l + d - 2)\, Y_{l,m}\left(\Omega\right)$ ($m=1,2,\dots, M$ and $M = \frac{(2l+d-2)\, (l+d-3)!}{l!\, (d-2)!}$ for $d=3,4,\dots$). Here $Z_{p,l}(\rho)$, with $p>0$ and $l=0,1,\dots$, solve the equation as follows:

\bqa
\left[\pr_\rho^2 + (d-1)\, \coth(\rho) \, \pr_\rho - \frac{l\, (l+d-2)}{\sinh^2(\rho)} \right]\, Z_{p,l} = - \left[p^2 + \left(\frac{d-1}{2}\right)^2\right]\, Z_{p,l}.
\eqa
They are given by

\bqa
Z_{p,l}(\rho) = \frac{\Gamma\left(ip + l + \frac{d-1}{2}\right)}{\Gamma\left(ip\right)}\,\sinh^{\frac{2-d}{2}}(\rho) \, P^{\frac{2-d}{2} - l}_{ip-\frac12}\left(\cosh\rho\right)
\eqa
and obey the orthogonality and completeness condition:

\bqa\label{orthonorm}
\int_0^\infty Z_{p,l}(\rho)\, Z^*_{p',l}(\rho) \, \sinh^{d-1}(\rho) \, d\rho = \delta\left(p-p'\right), \nonumber \\
\int_0^\infty Z_{p,l}(\rho)\, Z^*_{p,l}(\rho') \, dp = \sinh^{1-d}(\rho) \, \delta\left(\rho-\rho'\right)
\eqa
As the result we have the orthogonality and completeness condition for $H_{p,l,m}$

\bqa\label{ortcomcond}
\int d\rho\,d\Omega\, \sinh^{d-1}(\rho) \, H_{p',l',m'}\left(\rho,\Omega\right)\, H_{p,l,m}\left(\rho,\Omega\right) = \delta\left(p-p'\right)\, \delta_{ll'}\, \delta_{mm'},\nonumber \\
\int dp\, \sum_{l,m} \, H_{p,l,m}\left(\rho_1,\Omega_1\right)\, H_{p,l,m}\left(\rho_2,\Omega_2\right) = \frac{\delta\left(\rho_1-\rho_2\right)}{\sinh^{d-1}(\rho_1)}\, \delta\left(\Omega_1,\Omega_2\right)
\eqa
The convenient for our further discussion equality is as follows (see e.g. \cite{Grosche:1987de})

\bqa\label{equal}
\left(Z^2 - 1\right)^{\frac{2-d}{4}} \, \left|\frac{\Gamma\left(i\,p + \frac{d-1}{2}\right)}{\Gamma\left(i\, p\right)}\right|^2 \, P_{i\, p - \frac12}^{\frac{2-d}{2}}\left(Z\right) = \left(2\, \pi\right)^{\frac{d}{2}} \, \sum_{l,m} H^*_{p,l,m}\left(\rho_1, \Omega_1\right)\, H_{p,l,m}\left(\rho_2, \Omega_2\right),
\eqa
where $Z$ is the hyperbolic distance between $\left(\rho_1, \Omega_1\right)$ and $\left(\rho_2, \Omega_2\right)$; $P_a^b(Z)$ is the associated Legendre function. The useful for our discussion properties of these functions are: $P_a^b = P_{-a-1}^b$ and $P_a^m = \frac{\Gamma(a+m+1)}{\Gamma(a-m+1)}\, P_a^{-m}$ if $m \in Z$.

With the use of (\ref{ortcomcond}) and (\ref{equal}) one can prove that

\bqa
\int d\rho_2 d\Omega_2 \sinh^{d-1}(\rho_2) \,\left(Z_{12}^2 - 1\right)^{\frac{2-d}{4}} \, P_{i\, p_1 - \frac12}^{\frac{2-d}{2}}\left(Z_{12}\right) \times \nonumber \\ \times  \left(Z_{23}^2 - 1\right)^{\frac{2-d}{4}} \, P_{i\, p_2 - \frac12}^{\frac{2-d}{2}}\left(Z_{23}\right) \, \left|\frac{\Gamma\left(i\,p_2 + \frac{d-1}{2}\right)}{\Gamma\left(i\, p_2\right)}\right|^2 = \nonumber \\ = \left(2\, \pi\right)^{\frac{d}{2}}\, \delta\left(p_1 - p_2\right) \, \left(Z_{13}^2 - 1\right)^{\frac{2-d}{4}} \, P_{i\, p_1 - \frac12}^{\frac{2-d}{2}}\left(Z_{13}\right)
\eqa
More generally, using the orthogonality and completeness of the Legendre functions
(\ref{orthonorm}), with the substitution $\cosh \rho = Z$,
one can Mehler transform any function $F(Z)\to f(p)$ of the hyperbolic distance $Z$ \cite{Grosche:1987de}:

\begin{eqnarray}\label{Mehler}
F(Z)= \int_0^\infty \, f(p)\, P^{b}_{ip-1/2}(Z)
\, \left|\frac{\Gamma\left(\frac12 + ip - b\right)}{\Gamma\left(ip\right)}\right|^2 \, dp, \quad {\rm where}, \nonumber \\
f(p) = \int_1^\infty F(Z)\, P^{b}_{-ip-1/2}(Z) dZ
\end{eqnarray}
This transformation is well defined if $\int_1^\infty |F(Z)|^2\,dZ<\infty$ \cite{Vilenkin}. To obey the latter condition in the formulas below one has to apply Pauli--Villars UV regularization procedure.

In particular, the Mehler transform of the propagator in EAdS follows from (\ref{ortcomcond}) and (\ref{equal}):

\bqa\label{trans}
G(Z) = - \int_0^\infty dp \sum_{l,m} \frac{H_{p,l,m}\left(\rho_1,\Omega_1\right)\, H^*_{p,l,m}\left(\rho_2,\Omega_2\right)}{p^2 + \left(\frac{d-1}{2}\right)^2 + m^2} = \nonumber \\
= - \frac{\left(Z^2 - 1\right)^{\frac{2-d}{4}}}{\left(2\, \pi\right)^{\frac{d}{2}}} \, \int_0^\infty  \, \frac{P_{i\, p - \frac12}^{\frac{2-d}{2}}\left(Z\right)}{p^2 + \mu^2}\, \left|\frac{\Gamma\left(i\,p + \frac{d-1}{2}\right)}{\Gamma\left(i\, p\right)}\right|^2 \, dp.
\eqa
Now if $g(p)$ is the Mehler transform of $(Z^2-1)^{(d-2)/4}\, G_{reg}(Z)$ and $h(p)$ is the Mehler transform of $(Z^2-1)^{(d-2)/4}\, G^2_{reg}(Z)$, where $b=\frac{2-d}{2}$ in (\ref{Mehler}) and $G_{reg}$ is the Pauli--Villars regularized propagator, then the Mehler transform of the one loop contribution to the propagator, i.e. of $(Z^2-1)^{(d-2)/4}\, G_{1loop}(Z)$, is $(2\pi)^d \, \lambda^2 \, g^2(p)\, h(p)/4$.

Furthermore, using (\ref{orthonorm}) and (\ref{Mehler}) one can prove the Parseval equality $||F(Z)||^2 \equiv \int_1^\infty |F(Z)|^2 dZ = \int_0^\infty |f(p)|^2 \, \left|\frac{\Gamma\left(ip + \frac{d-1}{2}\right)}{\Gamma\left(ip\right)}\right|^2\, dp \equiv ||f(p)||^2$ \cite{Vilenkin}.
Then one has the following chain of relations $\frac{1}{(2\pi)^{2d}\, \lambda^4}\,||(Z^2-1)^{(d-2)/4}\,G_{1loop}(Z)|| = ||g^2(p)\, h(p)|| \le ||h(p)|| = ||(Z^2-1)^{(d-2)/4}\,G^2_{reg}(Z)|| < \infty$. The inequality follows from the fact that $||g^2(p)\, h(p)||^2 = \frac{1}{(2\,\pi)^{2d}}\,\int_0^\infty \frac{|h(p)|^2}{\left(p^2 + \mu^2\right)^4} \, \left|\frac{\Gamma\left(ip + \frac{d-1}{2}\right)}{\Gamma\left(ip\right)}\right|^2\, dp \le \int_0^\infty |h(p)|^2 \, \left|\frac{\Gamma\left(ip + \frac{d-1}{2}\right)}{\Gamma\left(ip\right)}\right|^2\, dp \equiv ||h(p)||^2$, because $|g(p)| = \frac{1}{(2\pi)^{d/2}(p^2 + \mu^2)}<1$. Thus, $G_{1loop}$ is finite and even decays fast as $Z\to \infty$ for the fields with any mass: even when $m=0$.

\section{Conclusions and Acknowledgments}

Loop corrections in the Poincare patch of dS space contain large IR contributions which behave as powers of
$\lambda^2\log(p\eta/\mu_{dS})$. Naive analytical continuation, $\mu_{dS} \to i\,\mu$ and $\eta \to i\, z$, to the Poincare patch of EAdS should produce contributions of the type $\lambda^2\log(pz/\mu)$. However, we show that such an analytical continuation is wrong. In particular, there are no large IR effects for the fields in Poincare patch of EAdS even for the massless fields. As well there are no IR divergences for the massless scalar fields in global EAdS. Thus, IR effects for non--conformal scalar fields in EAdS and AdS are even weaker than in Minkowski space. That should be expected on general physical grounds, because of the special confining EAdS geometry: The general physical discussion along these lines can be found in \cite{Callan:1989em}, while more rigorous tree level discussion is in \cite{Bros:2011vh}.

This observation points out that the large IR effects in dS space have a deeper physical reason rather than just its infinite space--time volume. In this paper we just make a small step towards clarification of the physical origin and consequences of the strong IR effects in dS space.

We would like to acknowledge discussions with A.Morozov, T.McLoughlin and A.Polyakov. AET would like to thank MPI, AEI, Golm, Germany for the hospitality during the final stage of the work on this project.
This work was supported by Grant "Leading
Scientific Schools" No. NSh-6260.2010.2, RFBR-11-02-01227-a and
Federal Special-Purpose Program "Cadres" of the Russian Ministry of
Science and Education. The work of ETA was partly supported by Ministry of Education and Science of the Russian Federation under the contract No. 14.740.11.0081. The work of SAV was partly supported by Ministry of Education and Science of the Russian Federation under the contract No. 14.740.11.0347 and by the Dynasty foundation.

\end{document}